\documentclass{article}
\usepackage{graphicx} 
\usepackage[utf8]{inputenc}

\usepackage{amsmath} 
\usepackage{amssymb}
\usepackage{amsthm}
\usepackage{xcolor} 
\usepackage{url}

\newtheorem*{thm*}{Theorem}
\newtheorem*{lemma*}{Lemma}
\newtheorem*{cor*}{Corollary}
\newtheorem*{prop*}{Proposition}

\theoremstyle{definition}

\title{  A Comprehensive  Study of Covid 19 in Florida
}
\author{ Julian Bennett \footnote{North Carolina State University, jabenne7@ncsu.edu }  and Lauren Eriksen \footnote{Purdue University, leriksen@purdue.edu }\\
Mentor Xingjie Helen Li \footnote{Department of Mathematics and Statistics, University of North Carolina at Charlotte,  xli47@uncc.edu} 
            }
\begin{document}
\maketitle

\begin{abstract}
Within the likes of any highly contagious and unpredictable disease, lies a predictable and attainable growth rate that researchers can find in order to make logistical conclusions about that particular disease and its affected regions. The foundation that researchers pull from when studying a particular disease and looking for its growth rate is the Susceptible-Infected-Removed (SIR) model, presented by a series of differential equations. The issue with the SIR model lies not in its complexity, but actually its simplicity and lack of a potentially high-finite amount of factors; the limit being bounded by the amount of data available for that particular factor. Our research involves the application of multiple regressions to pinpoint and identify Covid lockdown periods, followed by the modification of the SIR model. This involved creating new model approximations, such as the time-delayed SIR model and the reinfected SIR model, in order to take into account factors such as incubation and reinfection, and get the lowest error discrepancy as possible for our infection rate. We were able to conclude that as we included more factors, our error rate became lower and our results became more accurate. We could also identify outlier metropolitan areas and draw certain conclusions on performance level and the reasons behind them. We then moved on to find correlations, if any, between infection rates and outside factors. We looked at demographic and weather data to demonstrate whether correlations existed. We found that there are a few factors with high correlations, including graduate education and low temperatures.

\end{abstract}

\section{Introduction}



The rise of dangerous pandemics and epidemic viruses have left scientists and researchers with the need to counteract them. This could be anything from finding a way to measure them, how fast they spread, how long symptoms appear, the ability to get reinfected, and most importantly, a way to predict the spread of a particular disease. The SIR (susceptible-infected-removed) model was designed to be an epidemiological model for a disease that measures its particular growth rate \cite{Smith2004,cooper2020sir}.

SIR models are used very frequently to model disease infection rates, so we wanted to see how accurate it would be for Covid-19. However, attaining the most accurate prediction model is more complicated than it may seem, because these SIR equations don't account for many factors. While the original SIR model turns out to be too simplistic for states like Florida, it still remains a great foundational model. Throughout our research, we modified the original SIR equations to fit our data more closely. 

Our motivation behind the research came from using the SIR models for pandemic prediction. We wanted to see if it was possible to fit the parameters from data and accurately model infection rates for the disease. 

We can even consider a previous study of the spread of Covid-19 in North Carolina \cite{CovidNC2021}. This study was split into different metros that were dependent upon population size and popularity, and took into consideration the incubation rate and removal rate for Covid-19 victims. But even this modified SIR model for North Carolina does not fit our data for Florida. This then brings up other topics of interest like splitting up Florida metropolitan areas, finding the growth rate, and finding the incubation and removal rate of Covid-19. Furthermore, if these changes don't fit our data, we want to study the possible directions of modifying the original equations to get better predictive models of the spread of Covid in Florida.

We chose to study the Covid cases in Florida because this state is known for its tourism, education, politics, and overall popularity. These factors lead to large, diverse populations and visitors from lots of different places. We wanted to see if these factors had any affect on Covid infection rates.

In section 2, our research first explores what metropolitan areas we used and why, and in section 3, we explain what government parameters we used to split our data. We then will introduce the different SIR equations in section 4. In section 5, we explain linear regression and how we use it, and then follow up in section 6 with all our initial results. In section 7, our SIR equation will then be broken down into how and why these equations are relevant with our research relating to Covid in Florida, and what modifications we made to build upon the original SIR equation to improve the modeling. These results will be discussed in section 8. Lastly, in section 9, we will use multivariate regression to test for any correlation between Covid-19 cases and outside factors.


\section{Florida's Metro Areas}
In order to study different parts of Florida, we needed to find a way to combine counties because it was too hard to compare 68 different counties. While in total there are 22 metropolitan (metro) areas, we decided to have focus on the 11 major metro areas, based on population and popularity. These 11 metro areas all consisted of a population well over 500,000, while also being a very attractive tourism or political spot \cite{Metro}. Because our research hones in on the growth rate of Covid-19, these particular characteristics play a major role in our findings. Our 11 metro areas consist of Cape Coral, Deltona, Jacksonville, Lakeland, Miami, North Port, Orlando, Palm Bay, Pensacola, Tallahassee, and Tampa Bay. These metros covered the most important parts of Florida and made it easier to be able to compare areas. 

\section{Processing of Covid-19 Data in Florida}
Covid-19 was a worldwide epidemic that started back in March 2020. There were multiple government regulations to try to get Covid cases down so hospitals could stay afloat \cite{Covid-19Protocols}. We used these regulations from March 2020 to June 2020 to split up our data over time. This will help the quality of the linear regression model because the slopes and $R^2$ scores will better match the raw data. 

For example in March, there is a increase in Covid cases because very few Covid regulations had been enforced at that point. In April, however, there were many regulations enforced so there is a decrease in the rate of infection. These were the type of trends we were looking for when apply our model. Specially speaking, when we studied each metro, we would adjust the dates mentioned earlier in order to maximize the $R^2$ score for each time period. Once we found these dates over the entire time window, these became our lockdown periods for splitting and analyzing the data for each metro.

Meanwhile, these dates become more important when we compared metros because we were able to find when each metro actually followed protocols. This enabled us to identify the metros who were quick to follow these laws and who took a little longer. We then further investigated the reasons why particular metros were taking much longer to follow the protocols. In particular, we studied the factors that might have played a role in this, such as gender, education, location, and weather. These were the types of questions we pursued while conducting an in-depth analysis of the results from our research.

\section{Review of the original SIR Model}
When modeling and predicting the evolution of a pandemic, the simplest model to use is a susceptible-infected-recovered (SIR) model. It helps us find the growth rate of any particular disease, in our case Covid \cite{Smith2004,cooper2020sir}. This model aims to predict and estimate how people change from susceptible to infected to recovered over time. Mathematically, the SIR model is set up through a couple important variables, which are: 
\begin{itemize}
    \item $t$: number of passing days since first case of Covid;
            \item $S(t)$: number of susceptible people at time $t$;
            \item $I(t)$: number of infected people  at time $t$;
            \item $R(t)$: number of removed people  at time $t$;
            \item $\beta(t)$: average infection rate  at time $t$;
            \item $\gamma(t)$: average removal rate  at time $t$.
\end{itemize}
Often, one can assume constant parameters $\beta(t)\equiv \beta$ and $\gamma(t)\equiv \gamma$ to simplify the modeling in the first place.  
These variables can be put together into three important equations that will approximate the rate of change for each population in the SIR model.
\begin{subequations}\label{SIR}
\begin{equation} \label{eq1}
    \frac{dS(t)}{dt}= - \beta I(t)S(t),
\end{equation}
\begin{equation} \label{eq2}
    \frac{dI(t)}{dt}= \beta I(t)S(t) - \gamma I(t),
\end{equation}
 \begin{equation} \label{eq3}
    \frac{dR(t)}{dt}= \gamma I(t).
 \end{equation}
\end{subequations}

We are also given the initial points for each equation: $S(0)=S_0, \; I(0)=I_0,\;$ and $R(0)=R_0$. We aim to find the rate of change for susceptible, infected, and recovered populations from the data. Note that equation \eqref{eq1} is always negative. This is because everyone initially starts in the ``susceptible population'' and as people contract the disease, they will move to the ``infected population''. Equation \eqref{eq2} is the most important equation, though, because we want to fit the parameters from our data \cite{dong2022johns} and predict the growth rate of the Covid-19, that is $\frac{dI(t)}{dt}$.

Looking at equation \eqref{eq2} more closely, we can approximate this around $t=0$ (when the pandemic began). This will give us $\frac{dI(t)}{dt} \approx I(t)[\beta S_0 - \gamma]$. This equation has a solution of $I(t)=I(0) e^ {kt}$ where $k=\beta S(0) - \gamma$. This equation will be very important throughout our project because we can rearrange it to get $\log I(t) = kt + \log I_0$ and thus find the growth rate $k$ through linear regression over each lockdown period. 

On the other hand, if we add up all three equations, we can find that the right hand side adds up to zero, whereas the left hand side adds up to $\frac{d}{dt} ({S(t)+I(t)+R(t)})$. This means that $S(t)+I(t)+R(t)$ must equal a constant because the derivative of any constant equals zero. This tells us that the original SIR model always assumes a constant population, which is not the case in reality. In addition, the original SIR model does not take incubation period of Covid into account as all equations in \eqref{SIR} describe the same point in time: $t$. In later sections, we proposed modified SIR models to better model the reality of Covid-19 in Florida metros. 

\section{Study of Protocol Effectiveness}
Besides studying the growth rate of Covid-19 through the SIR model, we also studied the protocol effectiveness of each metro by using linear regression.
Linear regression 
is often shown in the form $y=\beta _0 +\beta _1 x$, where $\beta _1$ represents the slope and $\beta _0$ represents the $y$-intercept (bias). Connecting this concept to our project, we are using the data that we found for the number of infected people over time \cite{dong2022johns}, and graphing it in $\log$ scale domains. Recall that we divided the data over time where the initial division was based on the protocol release date.  We will then adjust the splitting dates per metro by checking the quality of the regression lines and $R^2$ scores. 
Once our regression lines are plotted and adjusted, we are able to pinpoint the actual date when each metro area started following protocol, which can be used as an assessment to study the protocol effectiveness.

\section{Initial Results}
For each metro area, we were able to create graphs using linear regression. We adjusted the five periods to represent our data in each graph, and found the best regression for each of these periods. One thing we wanted to find was when each metro started following protocols. This was found by adjusting the temporal length of each period and optimizing the quality of regression over that given period. 
We summarized all our finding into Table~\ref{Table1} so it was easier to compare each metro within Florida.

\begin{table}[htp!]
\begin{center}
\begin{tabular}{||c c c||} 
 \hline
 Metro & First Case & Protocol Followed  \\ [0.5ex] 
 \hline\hline
 Miami & 3/7 & 4/5 \\
 \hline
  Orlando & 3/12 & 4/5 \\
 \hline
 Lakeland & 3/17 & 4/7 \\
 \hline
Jacksonville & 3/11 & 4/8  \\
 \hline
 Tampa Bay & 3/2 & 4/8   \\
 \hline
 Tallahassee & 3/19 & 4/19  \\
 \hline 
 North Port & 3/8 & 4/23 \\
 \hline
 Deltona & 3/8 & 4/25 \\
 \hline 
 Cape Coral & 3/7 & 4/26  \\
 \hline
 Pensacola & 3/5 & 4/26 \\
 \hline 
  Palm Bay & 3/17 & 4/29   \\
 \hline
 \end{tabular}
  \end{center}
  \vspace{-0.5 cm}
 \caption{Study of protocol effectiveness in Florida. The actual protocol announcement date is March 29, 2020 \cite{Covid-19Protocols}.\label{Table1}}
\end{table}
When looking at this table, we can easily see the divide in metros that followed protocols early and those that took a lot longer. As shown in Table~\ref{Table1}, the first five metros obeyed the protocols much faster than the other six. In addition, these first five metros have some of the highest tourism in Florida. One would think that with more people, more cases would appear, but it actually turns out to be the other way according to this data. 

\section{Adjustments to the SIR Model}

In this section, we considered several approaches to enhance the original SIR equations \eqref{SIR}, taking into account the specific characteristics of Covid-19 and the evolving touristic features of Florida over time.

\subsection{Time-delayed SIR Model}
The first thing that needed to be done was adding a time-delay factor due to the incubation period of Covid. This means there is a certain amount of time that takes place between attracting the disease and symptoms appearing. In order to show this, we introduced $\tau_1$ into the original SIR model. $\tau_1$ gives us the average number of days for the incubation period. In addition, there is also a period of time for Covid-19 to be fully removed from the infected population, so we also introduced $\tau_2$ to take this into account. It gives us the average number of days for removal. With these variables, we created a new set of equations:
\begin{equation} \label{SIRdelay}
\begin{split}
    \frac{dS(t)}{dt}=& - \beta I(t-\tau _1)S(t),\\
    \frac{dI(t)}{dt}=& \beta I(t-\tau _1)S(t) - \gamma I(t-\tau _2),\\
    \frac{dR(t)}{dt}=& \gamma I(t-\tau _2).
    \end{split}
\end{equation}
We will use these equations as a basis to fit the parameters $\beta$ and $\gamma$ by comparing the growth rate $k$ from the data. 

 Since $k$ is the slope of original data, we can manipulate $\beta$ and $\gamma$ to match the slope of $k$ for each period. We have to divide all our data into 5 lockdown periods, so we are tuning $\gamma_1,\gamma_2,\gamma_3,\gamma_4,$ and $\gamma_5$ for each period by a visual comparison between the predicted solutions of the SIR model \eqref{SIRdelay} with the data on a log-scale domain. As a result, we will have a new $\beta$ which is the infection rate for each period. Although we tuned the data by a visual comparison, in a later section we introduce an error quantity to measure the discrepancy of slope $k$. 
 
\subsection{Repeated Infection SIR Model}
Another thing that has not been taken into account yet is the fact that people can get Covid multiple times. This means that after people enter the recovered population, they can later be re-entered into the susceptible population. This can be described by a new parameter $\mu$, which represents the fraction of people that survived their first Covid infection and have become susceptible again. Hence, the new set of equations read:
\begin{equation} \label{Srepeated}
\begin{split}
    \frac{dS(t)}{dt}&= - \beta I(t-\tau _1)S(t) + \mu R(t),\\
    \frac{dI(t)}{dt}&= \beta I(t-\tau _1)S(t) - \gamma I(t-\tau _2),\\
    \frac{dR(t)}{dt}&= \gamma I(t-\tau _2) - \mu R(t)
    \end{split}
\end{equation}
This version of the model is a more accurate representation of Covid-19 because many people ended up getting Covid multiple times. 

\subsection{New SIR Model with Incoming Tourism Flow}
Another way to improve the original SIR model \eqref{SIR}, is by accounting for the people that visited Florida for spring break, family, or work. Since Covid-19 started in March and many college students went to Florida for spring break, the state had tons of visitors that could have led to a significant increase in the spread of Covid. To take this factor into account, we introduce the parameter $\epsilon$ to account for the increase or decrease in the total population. This can be shown in this new set of equations:
\begin{equation} \label{SIR_tourism}
\begin{split}
    \frac{dS(t)}{dt}&= - \beta I(t-\tau _1)S(t) + \epsilon O(t),\\
    \frac{dI(t)}{dt}&= \beta I(t-\tau _1)S(t) - \gamma I(t-\tau _2),\\
    \frac{dR(t)}{dt}&= \gamma I(t-\tau _2), \\
    \end{split}
\end{equation}
$O(t)$ denotes the average incoming tourism population during the underlying period, which can be treated as an external source term.

Note that the only equation that changed was the first equation because the new people traveling into Florida are now seen as part of the susceptible population. The obstacle with this addition was finding the tourism data to account for people flying into the state. There is no overall data for each metro area that we currently considered. In addition, we were unsure how to account for people who drove into the state. Therefore, we will study and evaluate the performance of \eqref{SIR_tourism} in our future work.

\section{Results and Discussions}
\subsection{Numerical Approximations and Error Assessments}
We used the Euler method to numerically approximate the solutions of our proposed models. Because our time factor is counted on a daily basis, the time step size is fixed to $1$. For instance, the numerical approximation of \eqref{SIRdelay} becomes:
\begin{equation} \label{eq7}
\begin{split}
    \tilde{S}(t+1)&=\tilde{S}(t)-\beta \tilde I(t-\tau_1) \tilde S(t),\\
    \tilde I(t+1)&=\tilde I(t)+\beta \tilde I(t-\tau_1) \tilde S(t)-\gamma \tilde I(t-\tau_2),\\
    \tilde R(t+1)&=\tilde R(t)+\gamma \tilde I(t-\tau_2).
    \end{split}
\end{equation}
With these equations, we can input them into MATLAB and use the function dde23 to find values for $\tilde S(t)$, $\tilde I(t)$, and $\tilde R(t)$ at each discrete time grid $t$. The most important quantity of this output is the values of $\tilde I(t)$ because 
it represents the infected population. We can use these values to fit the growth rate over period $i$ of the SIR model by $\log \tilde I(t) = \tilde k_i t +\text{intercept}$. With this equation, we can find the slope $\tilde k_i$ over each period $i=1,\dots, 5$. This whole procedure was applied to the original SIR \eqref{SIR}, time-delayed SIR \eqref{SIRdelay} and repeated infection SIR \eqref{Srepeated}.

To mathematically compare the performance of various SIR models, we defined $|\tilde{k}_i-k_i|$ to compute the discrepancy between $\tilde{k}_i$ values from the SIR model with $k_i$ obtained from the data. In other words, it would show mathematically how similar our slopes were to each other.

However, because each lockdown period had a different length of time, we needed to take into account how long each period was when we computed $|\tilde{k}_i-k_i|$ for $i=1,\dots, 5$. 
We thus proposed a weighted error formula for a more accurate comparison. 
The new discrepancy formula that we came up with is:
\begin{equation} \label{discrepancy}
    \frac {\sum_{i=1}^5  |{\tilde{k_i} - k_i}| \times  \text {length of ith period}}{\text {total length of all periods}}.
\end{equation}
This definition of error quantity improved our comparison between the updated and original SIR models. 
With this formula, we were able to model the evolution of the pandemic within each metro over the five lockdown periods. 



\subsection{Comparison of Various SIR models}
With our time delayed and repeated infection SIR models, we tuned parameters and numerically computed the solutions to try to match our original data. We then calculated the errors of growth rates using our discrepancy formula \eqref{discrepancy} in each metro. 
We summarized our the errors in Table~\ref{Table2} to compare the performance.
\begin{table}[htp!]
    \begin{center}
    \begin{tabular}{||c c c||}
    \hline 
        Metro & Only Delayed & Reinfected \\
    \hline \hline 
        Cape Coral & 0.95\% & 0.66\% \\
    \hline 
        Deltona & 0.85\% & 0.36\% \\
    \hline 
        Jacksonville & 1.87\% & 1.53\% \\
    \hline 
        Lakeland & 0.67\% & 0.64\% \\
    \hline 
        Miami & 1.50\% & 0.48\% \\
    \hline 
        North Port & 1.83\% & 1.61\% \\
    \hline 
        Orlando & 1.86\% & 1.80\% \\
    \hline 
        Palm Bay & 0.90\% & 0.81\% \\
    \hline 
        Pensacola & 0.98\% & 0.91\% \\
    \hline 
        Tallahassee & 0.40\% & 0.27\% \\
    \hline 
        Tampa & 1.95\% & 1.66\% \\
    \hline 
    \end{tabular}
    \end{center}
    \vspace{-0.5 cm}
\caption{Error of growth rate of various modified SIR models. The error is defined in \eqref{discrepancy} and the reference growth rates are obtained from the raw data. \label{Table2}}
\end{table}

All of our errors were below 2\%, which means our model was quite accurate considering real world data. One big thing to point out, though, is that once we introduced this reinfected factor, the error rates for the repeated infection SIR model \eqref{Srepeated} decreased drastically for all metros. This means that the modified SIR became 
more accurate for modeling Covid spread when we included a time delay and the possibility of reinfection.  


\section{Statistical Correlation Study}
In order to find correlations between outside factors and Covid spread, we first have to introduce multivariate regression. This is just a slightly more complicated version of linear regression. Instead of having one dependent and one independent variable, multivariate regression contains one dependent and multiple independent variables. Each of these independent variables must also be independent of each other in order for the model to work correctly. When using multivariate regression, the main things to look for are the p-value and $R^2$ values. The p-value tells you if the data is significant or not, and the $R^2$ value tells you the correlation level. When looking at a p-value, you want to find if the value is above or below 0.05. If the number is above 0.05, the data is not significant and therefore the $R^2$ value is irrelevant. If the p-value is below 0.05, our data is significant and we can now look at the $R^2$ value. The $R^2$ value is a number between 0 and 1. The closer the number is to  1, the more correlated the data is and the better the model is. 

We wanted to set up two different multivariate regression equations. The first would include data from different age ranges, genders, income, and education. The second would include data from high and low temperatures, and different types of weather. Both of these factors were tested for correlation to Covid infection rates.

\subsection{Study of Demographic Data}
The first of the external categories of factors we researched was demographic data. We chose this as a correlation factor because things like age, gender, and education can play a significant role in the effects of certain situations. Also, demographic data is very accessible to the public, so it was easy to find. In this study, we modelled a multivariate regression that consisted of four independent variables: age, gender, income, and education \cite{MetroDemographic}. These were then split into multiple subcategories, as shown in the table below. All these variables depended upon our chosen $y$: the weighted average of Covid growth rate $k_i$. We chose this as our $y$ because the demographic data is given by the year, not by the day. We then had to normalize all the factors under one measurement so it could fit into one model. We chose to represent the factors in terms of percentages, as many were already shown this way in the data. 

\begin{table}[htp!]
\begin{tabular}{||c c||}
    \hline 
    Demographic & P-value  \\
    \hline \hline
    0-20 Age &  0.57568 \\
    \hline
    20-50 Age & 0.98726 \\
    \hline
    50-70 Age & 0.91047 \\
    \hline
    70-80+ Age & 0.74878 \\
    \hline
    Male & 0.31745 \\
    \hline
    Female & 0.22341  \\
    \hline
    0-50k Income & 0.37335 \\
    \hline
    50-100k Income & 0.36109  \\
    \hline
    100-150k Income & 0.37373  \\
    \hline
    150-200k$+$ Income & 0.33624  \\
    \hline
    High School Grad & 0.22308  \\
    \hline
    Some College & \color{purple}0.096412 \\
    \hline
    College Grad & 0.26149 \\
    \hline
    Post-Grad & \color{purple}0.042838 \\
    \hline 
    \end{tabular}
    \begin{tabular}{||c|c||}
    \hline \hline
     Demographic & $R^2$  \\
    \hline \hline 
    Age & 0.165 \\
    \hline
    Gender & 0.162 \\
    \hline
    Income & 0.516 \\
    \hline
    Education & \textcolor{purple}{0.833} \\
    \hline
    \end{tabular}
\caption{Multivariate regression study of Covid growth rate and demographic data. \label{Table3}}
    \end{table}

From Table~\ref{Table3}, the most observable trend is that the majority of the p-values seem to have no correlation to our Florida Covid-19 data, since almost all p-values are exceptionally above 0.05. For some factors, this is not surprising, however, for income this result proves very valuable in our research. In the previous year's research of Covid-19 in North Carolina, there was a correlation between education and income, but for Florida we only have a marginal correlation with education. Therefore, attaining a new result provides us with new conclusions that are not repetitive or redundant from the previous studies. 

The one factor that is significantly correlated with Covid cases in Florida is post-graduates. The 'Some College' factor is highlighted as well just for the sake of it being a significantly lower p-value than the other factors, even though it's not below 0.05. Looking back at the post-graduate demographics, we can interpret the statistical meaning behind its correlation, but need to look more closely at the $R^2$ before any conclusions are made.

When looking at $R^2$ values, we can only take the numbers into consideration if our p-values are below 0.05, otherwise they aren't significant. Since education had the only relevant p-value, we can see that its $R^2$ is relatively close to 1, showing that post-grad is an important factor for the regression modeling of Covid spread.

\subsection{Study of Weather Data}
Another one of the external factors we wanted to study was the different types of weather in each metro area. For each day, we found the highest temperature, the lowest temperature, and the type of weather (sunny, rainy, cloudy, foggy, snowy) in that specific metro \cite{Weather}. These represent our independent variables for the multivariate regression. Since these factors are all dependent on each other, we had to consider them separately. This first one we looked at was the type of weather. In order to handling these category data, we had to use dummy variables. 

Once we went through each metro, we found these results in Table~\ref{Table4}.
\begin{table}[htp!]
\begin{center}
    \begin{tabular}{||c c c c c c||}
    \hline 
    Metro & P1 & P2 & P3 & P4 & P5 \\
    \hline \hline
    Cape Coral & 0.661 & 0.892 & 0.567 & 0.732 & 0.727 \\
    \hline
    Deltona & 0.00112 & 0.678 & 0.914 & 0.636 & 0.524 \\
    \hline
    Jacksonville & 0.94 & 0.89 & 0.471 & 0.419 & 0.812 \\
    \hline
    Lakeland & NA & 0.763 & 0.58 & 0.401 & 0.0919 \\
    \hline
    Miami & 0.111 & 0.712 & 0.321 & 0.833 & 0.712 \\
    \hline
    North Port & 0.522 & 0.89 & 0.286 & 0.539 & 0.58 \\
    \hline
    Orlando & 0.468 & 0.966 & 0.669 & 0.0338 & 0.208 \\
    \hline
    Palm Bay & NA & NA & 0.362 & 0.683 & 0.371 \\
    \hline
    Pensacola & 0.363 & 0.974 & 0.551 & 0.526 & 0.393 \\
    \hline
    Tallahassee & NA & 0.165 & 0.392 & 0.112 & 0.593 \\
    \hline
    Tampa Bay & 0.11 & 0.772 & 0.529 & 0.347 & 0.573 \\
    \hline
    \end{tabular}
\end{center}
\vspace{-0.5 cm}
\caption{The p-values of multivariate regression of Covid growth rates over each lockdown period verse weather types for each metro. \label{Table4}}
\end{table}

From this table, we notice that the only p-values that were less than 0.05 were from Deltona period 1 and Orlando period 4. After looking at the corresponding $R^2$ values, we found that Deltona had a high value but Orlando had a much lower value. This means that our model was more accurate for Deltona, but not as much for Orlando.

After looking at our results from the different types of weather, we wanted to expand and look at the highest and lowest temperature for each day. We wanted to use the multivariate expression for these variables and compare to the data above to see if any new results appeared. For the highest values, we categorized temperatures above 80 as H (high), above 60 as M (medium), and anything below that as L (low). We then found the p-values for each metro and each of their periods in Table~\ref{Table5}.

\begin{table}[htp!]
\begin{center}
    \begin{tabular}{||c c c c c c||}
    \hline 
    Metro & P1 & P2 & P3 & P4 & P5 \\
    \hline \hline
    Cape Coral & 0.348 & NA & NA & 0.773 & NA \\
    \hline
    Deltona & 0.12 & 0.2 & 0.728 & 0.363 & 0.711 \\
    \hline
    Jacksonville & 0.422 & 0.0526 & 0.402 & 0.938 & NA \\
    \hline
    Lakeland & NA & 0.158 & 0.00389 & 0.153 & NA \\
    \hline
    Miami & 0.111 & NA & 0.0165 & 0.937 & NA \\
    \hline
    North Port & 0.315 & 0.00471 & 0.997 & 0.864 & NA \\
    \hline
    Orlando & NA & 0.72 & 0.0792 & 0.0248 & 0.505 \\
    \hline
    Palm Bay & NA & NA & 0.925 & 0.752 & 0.371 \\
    \hline
    Pensacola & 1 & 0.341 & 0.28 & 0.733 & NA \\
    \hline
    Tallahassee & NA & 0.338 & 0.218 & NA & 0.781 \\
    \hline
    Tampa Bay & 0.309 & 0.0924 & 0.836 & 0.972 & NA \\
    \hline
    \end{tabular}
\end{center}
\vspace{-0.5 cm}
\caption{The p-values of multivariate regression of Covid growth rates over each lockdown period verse daily highest temperatures for each metro. \label{Table5}}
\end{table}

From Table~\ref{Table5}, the most significant observation comes from Lakeland period 3, Miami period 3, North Port period 2, and Orlando period 4. All of these data points have a p-values of less than 0.05. From above, we notice that Orlando period 4 was also significant in the types of weather data. This means that we can look into this data and investigate causal factors of highest temperatures based on further studies. 

When we were looking at the lowest values, we categorized greater than 60 as H (high), greater than 50 as M (medium), and anything below that as L (low). We changed the cutoffs because we are looking at a totally different data set and our cutoffs need to reflect the numbers we have. Thus, the p-values are shown in Table~\ref{Table6}.

\begin{table}[htp!]
\begin{center}
    \begin{tabular}{||c c c c c c||}
    \hline 
    Metro & P1 & P2 & P3 & P4 & P5 \\
    \hline \hline
    Cape Coral & 0.786 & NA & 0.0673 & 0.228 & NA \\
    \hline
    Deltona & 0.332 & 0.0281 & 0.65 & 0.846 & NA \\
    \hline
    Jacksonville & NA & 0.0447 & 0.115 & 0.203 & NA \\
    \hline
    Lakeland & NA & 0.00056 & 0.307 & 0.174 & NA \\
    \hline
    Miami & NA & NA & NA & 0.927 & NA \\
    \hline
    North Port & 0.522 & 0.00019 & NA & 0.272 & NA \\
    \hline
    Orlando & NA & NA & 0.00000319 & 0.0104 & NA \\
    \hline
    Palm Bay & NA & NA & NA & NA & NA \\
    \hline
    Pensacola & 1 & 0.000000428 & 0.564 & 0.842 & NA \\
    \hline
    Tallahassee & NA & 0.329 & 0.758 & 0.397 & NA \\
    \hline
    Tampa Bay & 0.606 & 0.0000885 & NA & 0.094 & NA \\
    \hline
    \end{tabular}
\end{center}
\vspace{-0.5 cm}
\caption{The p-values of multivariate regression of Covid growth rates over each lockdown period verse daily lowest temperatures for each metro. \label{Table6}}
\end{table}

We can see that this table for lowest temperature looks a little different than the previous study based on highest temperatures. Here, we see a lot of zeros and quite a few extremely small values. The most important ones to look at are from Deltona period 2, Jacksonville period 2, Lakeland period 2, North Port period 2, Orlando period 3 and 4, Pensacola period 2, and Tampa Bay period 2. 

Evidently, the high and low temperatures have a better statistical correlation with Covid cases than the type of weather. Most of our data from the types of weather was inconclusive, but a decent amount of data from the high and low temperatures showed some correlation. 

\section{Conclusion}
Throughout our research, we have observed, compared, and analyzed 11 different metros in Florida. We initially found that about half of the metros followed protocols early, while the other half took about a month longer. Meanwhile, we studied the original and modified SIR models to model the spread of Covid-19 in Florida. After we adjusted our SIR model, we found that when we included the aspect of reinfection, our data become much more accurate. Lastly, we decided to find if any outside factors were correlated with the Covid infection rates in Florida. Through our research, we found that education and low temperatures had the most correlation. 

For the future, we would like to expand in a couple ways. Since we only focused on the first six months of 2020, we want to now look at the next six months to see if our model is still accurate in terms of prediction. In addition, we would like to look more deeply into certain correlations and what they really mean. We want to know why daily low temperatures and high education had a correlation with infection rates, and what this could mean for future pandemics or diseases.

\section*{Acknowledgements}
This work was supported in part by the NSF REU grant DMS-2150179.

    \bibliographystyle{abbrv}
    \bibliography{refs}

\end{document}